\documentclass[aps,twocolumn,showpacs]{revtex4}

\usepackage{graphics}
\usepackage{graphicx}
\def\bsigma{\mbox{\boldmath $\sigma$}}
\def\ss{\scriptscriptstyle }

\begin{document}
\title{Spin-dependent Rabi oscillations in single quantum
dot}

\author{F.T. Vasko$^*$}
\affiliation{NMRC, University College Cork, Lee Maltings
\\
Prospect Row, Cork, Ireland} 
\date{\today}

\begin{abstract}
Ultrafast optical pump of exciton resonance in single quantum dot by
an elliptically polarized laser pulse is described by the non-Markov 
balance equations.  Population and spin dynamics are investigated and 
spin-dependent Rabi oscillations are considered for the case of a flat
quantum dot, with the lateral size greater than the height. The following 
peculiarities of temporal evolution have been found: $a)$ a beating of 
Rabi oscillations under an elliptically polarized pump and $b)$ quenching 
of spin orientation due to a spin-orbit splitting effect.
\end{abstract}

\maketitle

Investigations of excitonic excitations in a single quantum dot (SQD) have 
been started in the past decade (see recent Refs. in \cite{1}). The 
Rabi oscillations and spin-related effects in a QD ensemble have also 
been investigated, see \cite{2} and \cite{3} respectively. Recently  the 
Rabi oscillations of electron population has been reported in
SQD excited by ultrashort laser pulse \cite{4,5}. Due to this, an examination 
of the spin-dependent ultrafast response in SQD is now timely. In this paper, 
we consider the spin-dependent coherent dynamics of an exciton in a SQD
under ultrafast interband pump by elliptically polarized radiation. 

A temporal evolution of the exciton is described by the $2\times 2$ 
spin-dependent density matrices written in the basis of the vacuum and excited 
excitonic states. These density matrices are governed by the balance equations 
with a non-Markov generation rate which is dependent on the population 
redistribution. For the simplest case of a weak spin-orbit interaction, 
allowing the only interband transitions between the same spin states, a + or 
$-$ circular polarized pump can only excite spin-down or spin-up states. Since
a temporal response under an elliptically polarized pump can be considered
as a superposition of Rabi oscillations under $\pm$ polarized pumps with 
different intensities, {\it a beating} of  Rabi oscillations takes place. 
For the more complicated case involving spin-orbit interaction in $c$-band, 
one 
can obtain {\it a quenching} of spin orientation in a SQD even under the 
circular polarized pump. Below, we have considered these phenomena in the 
framework of a flat QD model, with the lateral size greater than the height, 
and have estimated the maximal degree of spin orientation. 

We consider the coherent dynamics of a QD under resonant interband excitation
described by the perturbation operator $[\widehat{\delta H}_t\exp (-i\omega t)
+H.c.]$. The averaged over period, $2\pi /\omega$, density matrix, $\hat{R}_t$, 
obeys the following quantum kinetic equation (see evaluation in \cite{6,7}):
%1
\begin{eqnarray}
\frac{\partial \hat{R}_t}{\partial t}+\frac{i}{\hbar}[\hat{H},\hat{R}_t]=
\frac{1}{\hbar^2}\int_{-\infty }^{t}dt'e^{\delta t'} \\
\times\left\{ e^{i\omega (t-t')}[\hat{S}_{t-t'}[\widehat{\delta H}_{t'},
\hat{R}_{t'}]\hat{S}_{t-t'}^+,\widehat{\delta H}_t^+] +H.c.
\right\} , \nonumber
\end{eqnarray}
where $\delta\rightarrow +0$ and $\hat{S}_t=\exp (i\hat{H}t/\hbar )$
describes evolution of QD with the Hamiltonian $\hat{H}$. The transitions 
between 
the spin-degenerate $c$-band and $v$-band states are described by 
the matrix
%2
\begin{equation}
\widehat{\delta H}_t=\frac{ie}{\omega}Ew_t\left|\begin{array}{ll} 0 & 
\hat{g}_{\bf e} \\ \hat{g}_{\bf e}^+ & 0 \end{array}\right| .
\end{equation}
Here we consider the interband photogeneration caused by the electric field 
$E[{\bf e}\exp (-i\omega t)+c.c.]w_t$,  with the unit complex polarization 
vector ${\bf e}$, and $w_{t}=\exp [-(t/\tau_{p})^2]$ is the Gaussian 
form-factor of the pulse, $\tau_p$ stands for the pulse duration. The 
polarization-dependent matrix $\hat{g}_{\bf e}=({\bf e}\cdot\hat{\bf v}_{cv})$
is written through the interband velocity matrix element, $\hat{\bf v}_{cv}$, 
and $\hat{g}_{\bf e}$ is the non-Hermitian matrix because ${\bf e}\neq
{\bf e}^*$.

Below, we shall restrict our consideration to the vicinity of the excitonic 
resonance when $\hat{R}_t$ can be replaced by the diagonal matrix $\left|
\begin{array}{ll} \hat{\rho}_{e,t} & 0 \\ 0 & \hat{\rho}_{o,t}\end{array}
\right|$ with $2\times 2$ matrices $\hat{\rho}_{e,t}$ and $\hat{\rho}_{o,t}$ 
describing the excited ($e$-) and vacuum ($o$-) states respectively. In 
agreement with the experimental conditions \cite{5}, we have neglected here 
the two-photon biexcitonic resonance, see \cite{2,8} and Refs. therein. Thus, 
we transform Eqs.(1,2) into the following form:
%3
\begin{eqnarray}
\frac{\partial}{\partial t}\left|\begin{array}{l} \hat{\rho}_{e,t} \\ 
\hat{\rho}_{o,t}\end{array}\right| =\pm\left(\frac{eE}{\hbar\omega}
\right)^2w_t\int_{-\infty }^{t}dt'w_{t'}e^{i\Delta\omega (t'-t)} \nonumber \\
\times\left( \hat{g}_{\bf e}\hat{\rho}_{o,t'}-\hat{\rho}_{e,t'}
\hat{g}_{\bf e}\right)\hat{g}_{\bf e}^+ +H.c., 
\end{eqnarray}
where $\hbar\Delta\omega$ is the detuning energy with respect to the exciton 
peak energy. Adding these Eqs. we obtain that $\hat{\rho}_{et}+\hat{\rho}_{ot}$
is a time-independent matrix which is determined through initial conditions
(see below).

To calculate the matrix $\hat{g}_{\bf e}$ we neglect the Coulomb
 renormalization
of the interband velocity matrix element and consider
%4
\begin{equation}
\langle\lambda |\hat{\bf v}|\lambda '\rangle\simeq \sum_{jj'}{\bf v}_{jj'}
\int d{\bf x}\psi^{\lambda~*}_{j{\bf x}}\psi^{\lambda '}_{j'{\bf x}} .
\end{equation}
Here ${\bf v}_{jj'}$ is the $8\times 8$ interband velocity matrix and 
$\bf x$ is the in-plane vector. For the flat QD model, we write the overlap 
integral in (3) through the in-plane envelope functions $\psi^{\lambda}_{j
{\bf x}}$. Here we have replaced the transverse contribution to the overlap 
integral by unit considering transitions between top $v$-band states and 
bottom $c$-band states. Neglecting the nonparabolic ($\propto p^3$) 
contributions, we describe the $v$-band states by $\varphi^{\ss (h)}_{\bf x}
|\sigma\rangle$, where $\varphi^{\ss (h)}_{\bf x}$ takes into account the 
heavy-light hole mix \cite{6} and $|\sigma\rangle$ corresponds to the 
degenerate spin state directed along $OZ$ with the spin numbers $\sigma =
\pm 1$. Taking into account noticeable spin-orbit effects in $c$-band we 
write the eigenstate problem for spinor $\psi_{\bf x}^{\ss (c)}$ as follows:
%5
\begin{equation}
\left\{\frac{\hat{p}^2}{2m}+(\hat{\bsigma}\cdot[{\bf v}_s\times\hat{\bf p}])
-\varepsilon\right\} \psi_{\bf x}^{\ss (c)}=0 ,
\end{equation}
where $\hat{\bsigma}$ are the Pauli matrices, $m$ is the electron effective mass, and 
the characteristic spin velocity ${\bf v}_s$ is along the growth axis, $OZ$ 
\cite{6,9}. 

We shall restrict our consideration to the case of weak spin-orbit effect,
performing the unitary transformation $\psi_{\bf x}^{\ss (c)}=\exp [-im(
\bsigma\cdot [{\bf v}_s\times{\bf x}])/\hbar ]\varphi_{\bf x}$. Taking into 
account $\propto{\rm v}_s^2$ contributions only, we obtain the Schrodinger 
equation in the form:
%6
\begin{equation}
\left\{\frac{\hat{p}^2}{2m}-2m{\rm v}_s^2[1-(\hat{\bsigma}\cdot [\hat{\bf p}
\times{\bf x}])/\hbar ]-\varepsilon\right\}\varphi_{\bf x} =0 .
\end{equation}
We consider this equation with the zero boundary condition $\varphi_{{\bf x}|
\it{\Gamma}}^{\ss (c)}=0$; the curve $\it{\Gamma}$ is bounded by a flat QD with the area 
$S$.
 Note that the momentum operator here, $[\hat{\bf p}\times{\bf x}]/\hbar$,
is along $OZ$, so that the spinor $\varphi_{\bf x}$ is proportional to the
above introduced spinor $|\sigma\rangle$.

Next, using the two-level basis including the lower electron states with $\sigma
=\pm 1$ and the upper hole states with $\sigma '=\pm 1$, we obtain the only
non-zero matrix elements (2) for the transitions $|e,1\rangle\rightarrow 
|o,1\rangle$ and $|e,-1\rangle\rightarrow |o,-1\rangle$; see similar selection 
rules in \cite{6,10}. For the case of elliptically polarized excitation 
described by ${\bf e}=\lambda_+{\bf e}_+ +\lambda_-{\bf e}_-$, where 
${\bf e}_{\pm}$ are the polarization vectors for $\pm$ circular waves and 
$\lambda_{\pm}$ are complex numbers (moreover $|\lambda_+|^2+|\lambda_-|^2=1$),
we obtain the matrix $\hat{g}_{\bf e}$ in the form:
%7
\begin{equation}
\hat{g}_{\bf e}={\rm v}_{\ss\bot}\left|\begin{array}{ll} \lambda_- & 0 \\
0 & -\lambda_+ \end{array}\right| .
\end{equation}
The non-zero coefficients here are written through
%8
\begin{equation}
{\rm v}_{\ss\bot}=\int d{\bf x}\psi^{\ss (c)}_{\bf x}\varphi^{\ss(h)}_{\bf x}
\simeq{\cal P}\left[ 1-\alpha\left(\frac{m{\rm v}_s}{\pi\hbar}
\right)^2S\right] , 
\end{equation}
where ${\cal P}$ is the Kane velocity. The overlap integral here have been 
calculated for the cases of circular and square QDs when the numerical 
coefficient $\alpha$ is equal
 to 0.11 and 0.32 respectively. A suppression of 
the matrix element (7) with 
increasing ${\rm v}_s$ also takes place for the 
conic QD case \cite{11}.

We describe the coherent dynamics by the use of the diagonal population
numbers for the excited and ground states, $n_{e,\sigma t}=\langle e,\sigma |
\hat{\rho}_{e,t}|e,\sigma \rangle$ and $n_{o,\sigma t}=\langle o,\sigma |
\hat{\rho}_{o,t}|o,\sigma \rangle$. The system (3) is transformed into the 
independent equations for the states with $\sigma =\pm 1$:
%9
\begin{eqnarray}
\frac{d}{dt}\left|\begin{array}{l} n_{e,{\ss\pm 1}t} \\ n_{o,{\ss\pm 1}t}
\end{array}\right|=\pm\nu_r|\lambda_{\mp}|^2 \\
\times\int_{-\infty }^{t}\frac
{dt'}{\tau_p}\Phi_{tt'}\left( n_{o,{\ss\pm 1}
t'}-n_{e,{\ss\pm 1}t'}\right) .
 \nonumber
\end{eqnarray}
Here we have introduced the kernel $\Phi_{tt'}=w_tw_{t'}\cos\Delta\omega 
(t-t')$, and $\nu_r=2(eE/\hbar\omega )^2|{\rm v}_{\ss\bot}|^2\tau_p$ is the 
characteristic photogeneration rate. Since $n_{e,{\ss\pm 1}t}+n_{o,{\ss\pm 1}
t}$ are independent of time, and the oscillation strengths for
$\pm$ transitions are proportional to $|\lambda_{\mp}|^2$, we use the 
phenomenological conditions in the form 
%10
\begin{equation}
n_{e,{\ss\pm 1}t}+n_{o,{\ss\pm 1}t}
=|\lambda_{\mp}|^2,
\end{equation}
which is in agreement with the particle conservation requirement $\sum_{\sigma}
(n_{e\sigma t}+n_{o\sigma t})=1$. Using Eqs.(9), we obtain 
the closed 
equation for the population differences, $\Delta n_{\pm t}=n_{o,
{\ss\pm 1}t}-
n_{e,{\ss\pm 1}t}$, in the form:
%11
\begin{equation}
\frac{d\Delta n_{\pm t}}{dt}+\nu_r|\lambda_{\mp}|^2\int_{-\infty }^{t}\frac
{dt'}{\tau_p}\Phi_{tt'}\Delta n_{\pm t}=0 
\end{equation}
with the initial conditions $\Delta n_{\pm t\rightarrow -\infty}=|
\lambda_{\mp}|^2$ which are obtained from Eq.(10) and $n_{e,{\ss\pm 1}t
\rightarrow -\infty}=0$.

The concentration of photoexcited electrons is given by $n_t\equiv n_{e,{\ss 1}
t}+n_{e,{\ss -1}t}$ while the spin orientation, ${\bf S}_t$, is determined as 
follows ${\bf S}_t=\sum_{\sigma =\pm 1}\langle e,\sigma |\hat{\bsigma}|e,
\sigma\rangle n_{e,\sigma t}$. The only $z$-component of spin appears to be 
non-zero and 
%112
\begin{eqnarray}
{\rm S}_{et}^{z}=k(n_{e,{\ss 1}t}-n_{e,{\ss -1}t})\equiv k\delta n_t, 
\nonumber \\
k\simeq 1-2\alpha\left(\frac{m{\rm v}_s}{\pi\hbar}\right)^2S ,
\end{eqnarray}
where $k=\int d{\bf x}\psi^{{\ss (c)}+}_{\bf x}\hat{\sigma}_z\psi^{\ss (c)}
_{\bf x}$ and $\alpha$ is the coefficient from Eq.(8). The spin orientation
is proportional to the redistribution between the $\pm$ spin states,
$\delta n_{t}$, and the factor $k$ decreases with increasing ${\rm v}_s$. 

For the resonant excitation case, $\Delta\omega =0$, the kernel in Eq.(11) is 
not dependent on $t-t'$, one obtains the second-order differential equation
%13
\begin{equation}
\frac{d^2\Delta n_{{\ss\pm}t}}{dt^2}-\frac{2t}{\tau_p^2}\frac{d\Delta 
n_{{\ss\pm}t}}{dt}+\frac{\nu_r}{\tau_p}|\lambda_{\mp}|^2E^{-2(t/\tau_p)^2}
\Delta n_{{\ss\pm}t}=0 
\end{equation}
with the use of an additional initial condition $[w_t^{-1}d\Delta n_{{\ss\pm}t}
/dt]_{t\rightarrow -\infty}=0$. The solution of Eq.(13) takes the form:
%14
\begin{equation}
\Delta n_{{\ss\pm}t}=|\lambda_{\mp}|^2\cos\left(\frac{A|\lambda_{\mp}|}
{\sqrt{\pi}}\int_{-\infty}^{t/\tau_p}d\tau e^{-\tau^2}\right) , 
\end{equation}
where $A=\sqrt{\pi\nu_r\tau_p}$ is the excitation amplitude and $|\lambda_{\pm}
|^2=(1\pm s)/2$ with the degree of circular polarization $s=|\lambda_+|^2-|
\lambda_-|^2$. 

In Fig.1 we plot the temporal evolution of $n_t$ and $\delta n_t$ for the cases
of circular, elliptical (with $s=1/2$), and linear polarized pumps under the
resonant excitation condition $\Delta\omega =0$. Note that $\delta n_{t}=
-n_{t}$ and $\delta n_{t}=0$ for the cases of circular and linear polarized 
pumps 
respectively. For the $\pi$-pulse excitation case (if $A=\pi$), one 
obtains monotonic increase of $n_t$ with $n_{t\rightarrow\infty}=1$ for the 
circular 
pump. The spin redistribution, $\delta n_{t}<0$, varies with $t/\tau_p$
in a 
similar manner. As the excitation amplitude increases, the temporal 
evolution 
becomes oscillatory, as shown in Fig.1$b$ for $A=3.5\pi$. These 
dependencies have different periods, for the circular, linear, and elliptically
polarized pumps.
\begin{figure}
\begin{center}
\includegraphics{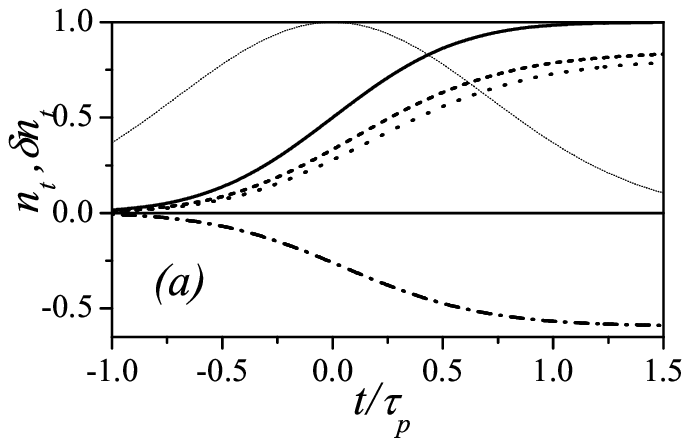}
\includegraphics{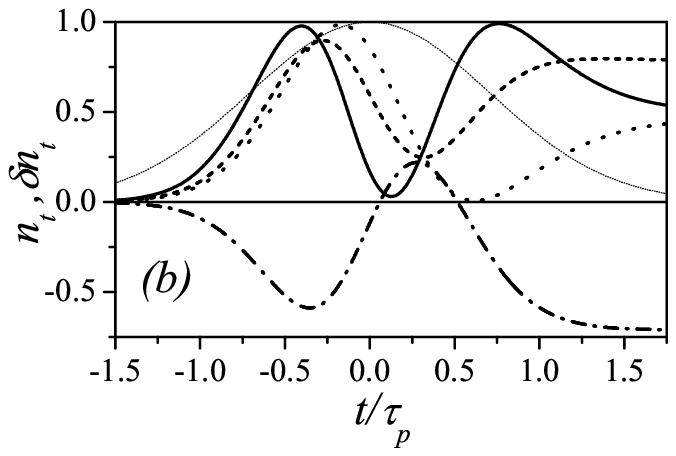}
\end{center}
\addvspace{-1 cm}\caption{Temporal evolution of the total population, 
$n_t$, and the spin redistribution, $\delta n_{t}$, under the excitation 
amplitudes, $A=\pi$ ($a$) and $A=3.5\pi$ ($b$). Solid, dashed and dotted 
curves correspond $n_t$ under circular, elliptical and linear polarized 
pumps respectively. The dot-dashed curve presents $\delta n_{t}$  under 
elliptical polarized pump. The pulse profile is shown as the short-dot 
curve.}
\label{fig.1}
\end{figure}

Next we consider the detuning effect by calculating the solution of
 Eq.(11)
numerically. In Fig.2 we plot the temporal evolution of $n_{t}$ and $\delta 
n_{t}$
 under different $\Delta\omega\tau_p$. The detuning results in 
suppression 
of both the Rabi flop after the pulse and the amplitude of 
oscillations if $A>\pi$.
 The characteristics of the temporal response for 
different polarizations of pump do not 
change in comparison with the resonant 
case.
\begin{figure}
\begin{center}
\includegraphics{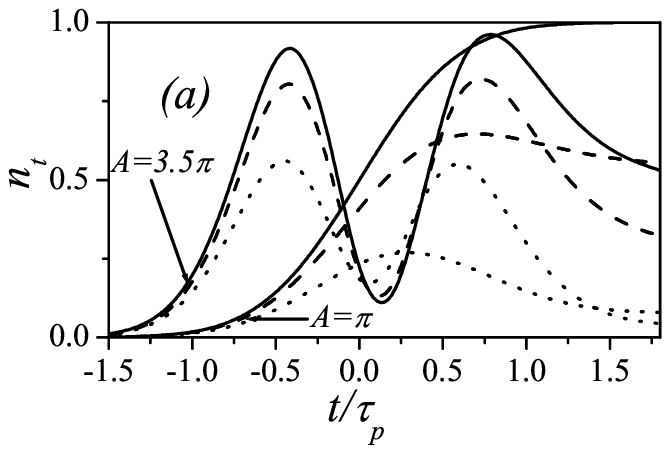}
\includegraphics{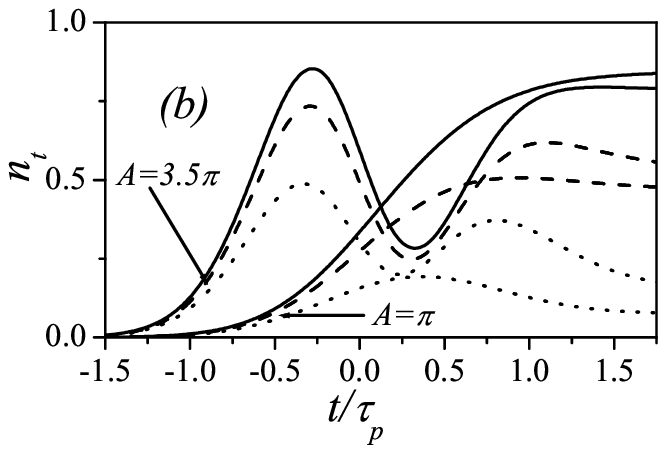}
\includegraphics{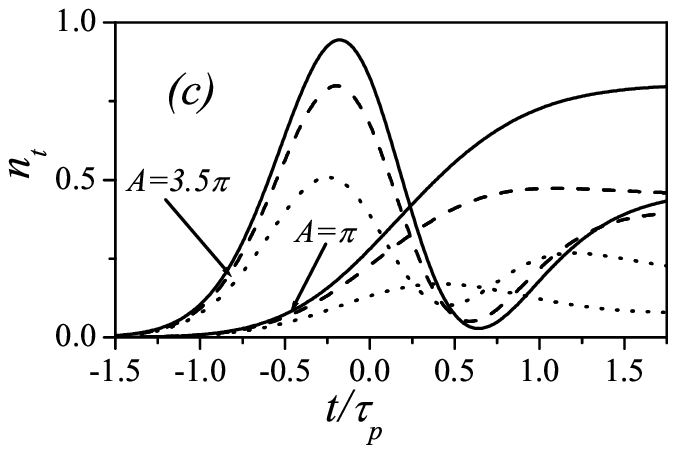}
\includegraphics{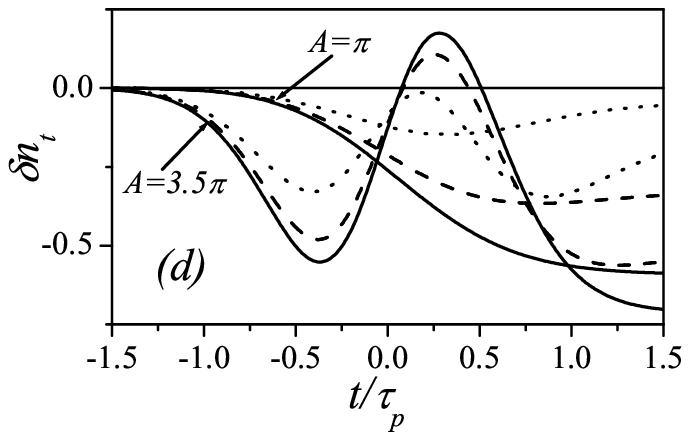}
\end{center}
\addvspace{-1 cm}\caption{Quencing of the temporal response, both the total 
population $n_t$ ($a$-$c$) and the spin redistribution $\delta n_t$ ($d$),
under resonant (solid curves) and non-resonant, with $\Delta\omega\tau_p =\pi 
/3$ (dashed curves) and $\Delta\omega\tau_p =2\pi /3$ (dotted curves),
pumps. The excitation amplitudes, $A$, are indicated by arrows.}
\label{fig.2}
\end{figure}

The oscillations of Rabi flopping population, $n\equiv n_{t\rightarrow\infty}$,
and the redistribution between the $\pm$ spin states, $\delta n\equiv\delta 
n_{t\rightarrow\infty}$, are shown in Fig.3 for the resonant excitation case. 
The harmonic dependencies with periods $A=2\pi$ and $A=2\sqrt{2}\pi$ 
take place for the circular and linear polarized pumps respectively. By
contrast, one can see a beating of $n$ and $\delta n$ with $A$ under the 
elliptically polarized pump. Moreover, the spin orientation may change a sign 
for this case; once again, $\delta n=-n$ and $\delta n=0$ for the circular 
and linear polarized cases respectively. 
\begin{figure}
\begin{center}
\includegraphics{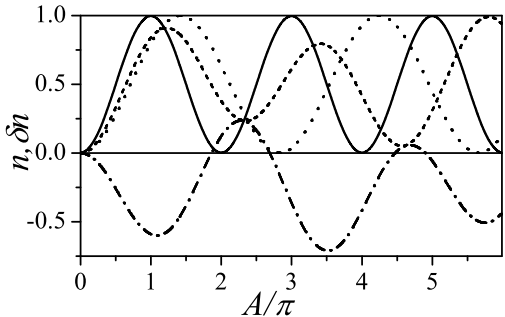}
\end{center}
\addvspace{-1 cm}\caption{Oscillations of Rabi flopping population in $c$-band, 
$n$, and the spin redistribution, $\delta n$, versus excitation amplitude, 
$A$ (the same curves as in Fig.1).}
\label{fig.3}
\end{figure}

Let us discuss the assumptions used in the above calculations. The main
restriction is the consideration of the vicinity of the excitonic resonance
without the contribution of two-photon biexcitonic resonance. Here we have 
applied the phenomenological initial conditions for Eq.(11). This 
is valid if $\hbar\Delta\omega$ and $\hbar /\tau_p$ are smaller than the 
exciton-biexciton splitting energy (around 3meV in $InGaAs$ QDs) while a more 
detailed many-particle consideration is required for greater 
$\hbar\Delta\omega$ and/or $\hbar /\tau_p$. We also have described the coherent
response neglecting
relaxation because $\tau_p$ is shorter than the relaxation times. Next, the 
non-Markov equation (1) describes the interband generation of electron-hole 
pairs to second order accuracy, i.e. we have supposed that the energy 
$|e|E{\rm v}_{\ss\bot}/\omega$ is much lower than the energy gap. Finally, the 
simple model of QD has been used in order to obtain Eqs.(8,12) and more 
complicated numerical calculations, e.g. similar to \cite{12}, have to be 
performed in order to improve these simple formulas.

In conclusion, we have considered the peculiarities of Rabi oscillations in 
a single quantum dot due to the spin-dependent ultrafast excitation by an 
elliptically polarized pulse. For the simple model of SQD with the spin-flip 
transitions forbidden we have found both the beating of Rabi oscillations 
under an elliptically polarized pump and the quenching of spin orientation 
due to spin-orbit splitting effect. A more detailed treatment of these effects 
requires the consideration of complicated selection rules and to describe the
exciton-biexciton resonance in the framework of a many-particle approach.
What this paper does, however, is to determine the conditions for 
observation of the peculiarities mentioned.

\textbf{Acknowledgment}:
This work was supported by Science Foundation Ireland. I would like to
thank E.P. O'Reilly, B. Roycroft, and R. Zimmermann for useful discussions.

$^{*}$ E-mail: {\rm fvasko@yahoo.com}. 

On leave from: Institute of Semiconductor Physics, Kiev, 03650, Ukraine.


\begin{references}
\bibitem{11}
The description of the conic QD is somewhat complicate. For the $In_{0.5}
Ga_{0.5}As$ QD with height, $h$, radius, $R$, and the area, $S=\pi R^2$, the
parameter $\alpha$ in Eq.(8) is replaced by $1.3(h/R)^{4/3}$ and $\cal P$
is reduced by $\simeq 10\%$. At the same time, the factor $2\alpha$ in Eq.(12)
is replaced by $5.9(h/R)^{4/3}$, i.e. the spin orientation is stronger
suppressed.
\end{references}
\end{document}